\documentclass[12pt]{iopart}

\usepackage{graphicx}

\begin{document}

\title[Simulation of the thermal expansion of a plasma]{Simulation of a 
collision-less planar electrostatic shock in a proton-electron plasma 
with a strong initial thermal pressure change}

\author{M E Dieckmann\textsuperscript{1,2,3},G Sarri\textsuperscript{1},
L Romagnani\textsuperscript{1}, I Kourakis\textsuperscript{1} and 
M Borghesi\textsuperscript{1}}

\address{1 Centre for Plasma Physics, Queen's University Belfast, Belfast
BT7 1NN, U K}
\address{2 Theoretische Physik IV, Ruhr-University Bochum, 44780 Bochum,
Germany}
\address{3 Department of Science and Technology (ITN), Link\"oping 
University, Campus Norrk\"oping, 60174 Norrk\"oping, Sweden}
\ead{Mark.E.Dieckmann@itn.liu.se}

\begin{abstract}
The localized deposition of the energy of a laser pulse, as it ablates
a solid target, introduces high thermal pressure gradients in the plasma. 
The thermal expansion of this laser-heated plasma into the ambient medium 
(ionized residual gas) triggers the formation of non-linear structures in 
the collision-less plasma. Here an electron-proton plasma is modelled with 
a particle-in-cell (PIC) simulation to reproduce aspects of this plasma 
expansion. A jump is introduced in the thermal pressure of the plasma, 
across which the otherwise spatially uniform temperature and density change 
by the factor 100. The electrons from the hot plasma expand into the cool 
one and the charge imbalance drags a beam of cool electrons into the hot 
plasma. This double layer reduces the electron temperature gradient. The 
presence of the low-pressure plasma modifies the proton dynamics compared 
to the plasma expansion into a vacuum. The jump in the thermal pressure 
develops into a primary shock. The fast protons, which move from the hot 
into the cold plasma in form of a beam, give rise to the formation of phase 
space holes in the electron and proton distributions. The proton phase space 
holes develop into a secondary shock that thermalizes the beam.
\end{abstract}

\pacs{52.38.Hb,52.35.Qz,52.65.Rr}
%\no

\maketitle

\section{Introduction}

The impact of a laser pulse on a solid target results in the evaporation 
of the target material. The heated plasma expands under its own thermal 
pressure and shocks as well as other nonlinear plasma structures form. 
Generating collision-less plasma shocks in a laboratory experiment 
permits us to study their detailed dynamics in a controlled manner. A 
better understanding of such shocks is not only relevant for the laser-plasma 
experiment as such and for inertial confinement fusion experiments. It can
also provide further insight into the dynamics of solar system shocks and 
the nonrelativistic astrophysical shocks, like the supernova remnant 
shocks \cite{Koopman,Bell,Remington,Woolsey,Drury}. 

An obstacle to an in-depth investigation of the laser-generated shocks has 
been so far, that the frequently used optical probing techniques could not 
resolve the shock structure at the required spatio-temporal resolution. 
The now available proton imaging technique \cite{B1,B2} helps us overcoming 
this limitation. This method can provide accurate spatial electric field 
profiles at a high time resolution, as long as no strong magnetic fields 
are present. The nonrelativistic flow speed of the laser-generated shock, 
e.g. that in Ref. \cite{Romagnani}, implies that no strong self-induced 
magnetic fields due to the filamentation instability or the mixed mode 
instability \cite{Filament,Bret} occur at the shock front. 

The availability of electric field data at a high resolution serves as
a motivation to perform related numerical simulations and to compare
their results with the experimental ones. The experimental observations 
from Ref. \cite{Romagnani}, which are most relevant for the simulation 
study we perform here, can be summarized as follows. The ablation of a 
solid target consisting of aluminium or tungsten by a laser pulse with a 
duration of $\approx$ 470 ps and an intensity of $10^{15}$ W/${\rm cm}^2$ 
results in a plasma with a density $\approx 10^{18}$ ${\rm cm}^{-3}$ and with 
an electron temperature of a few keV. This plasma expands into an ambient 
plasma with the density $\leq 10^{15}$ ${\rm cm}^{-3}$. The ambient plasma has 
been produced mainly by a photo-ionization of the residual gas. The dominant 
components of the residual gas, which consists of diluted air, are oxygen 
and nitrogen. Electrostatic structures, which move through the ionized 
residual gas, are observed. Their propagation speeds suggest that one is 
an electrostatic shock \cite{Forslund} with a thickness of a few electron 
Debye lengths, which is expanding approximately with the ion acoustic 
velocity $2-4 \times 10^{5}$ m/s. Ion-acoustic solitons are trailing the 
shock. Another structure moves at twice the shock speed, which is probably 
related to a shock-reflected ion beam. The electron-electron, electron-ion 
and ion-ion mean free paths for the residual gas have been determined for 
this particular experiment. They are of the order of cm and much larger 
than the shock width of a few tens of $\mu$m. The shock and the electrostatic 
structures are collision-less.

The experiment can measure the electric fields, the propagation speed of 
the electric field structures and it can estimate the electron temperature 
and density. The bulk parameters of the ions, such as their temperature, 
mean speed and ionization state, are currently inaccessible, as well as 
detailed information about the spatial distribution of the plasma. We can 
set up a plasma simulation with the experimentally known parameters, and 
we can introduce an idealized model for the unknown initial conditions. 
The detailed information about the state of the plasma, which is provided 
by Vlasov simulations \cite{Cheng} or by particle-in-cell (PIC) simulations 
\cite{Dawson,Eastwood}, can then provide further insight into the expansion 
of this plasma. 

Here we investigate a mechanism that could result in the shock observed 
in Ref. \cite{Romagnani}. We model with PIC simulations the interplay of two 
plasmas with a large difference in the thermal pressure, which are 
initially spatially separated. We aim at determining the spatio-temporal 
scale, over which a shock forms under this initial assumption, and we 
want to reveal the structures that develop in the wake of the shock. 
The temperature and density of the hot laser-ablated plasma both exceed 
initially that of the cold ambient plasma by two orders of magnitude. 
The density ratio is less than that between the expanding and the 
ambient plasma in Ref. \cite{Romagnani}. However, the density will not 
change in form of a single jump in the experiment and realistic density 
changes will probably be less or equal to the one we employ. Selecting 
the same jump in the density and temperature is computationally efficient, 
because both plasmas have the same Debye length that determines the grid 
cell size and the allowed time step. The ion temperature in the experiment 
is likely to be less than that of the electrons. The electron distribution 
can also not be approximated by two separate spatially uniform and thermal 
electron clouds, because the plasma generation is not fast compared to the 
electron diffusion. We show, however, that the shock forms long after the 
electrons have diffused in the simulation box and reached almost the same 
temperature everywhere. 

A change in the thermal pressure by a factor $10^4$ should imply a plasma 
expansion that is similar to that into a vacuum. This process has received 
attention in the context of auroral, astrophysical and laser-generated 
plasmas and it has been investigated analytically within the framework of 
fluid models \cite{SackRev,Sack} or Vlasov models \cite{Manfredi,Dorozhkina}. 
It has been modelled numerically using a cold ion fluid and 
Boltzmann-distributed electrons \cite{Mora} and with kinetic Vlasov and PIC 
simulations \cite{Mora3,Mora2}. The plasma expansion of hot electrons and 
cool ions into a tenuous medium has also been examined with PIC simulations, 
such as the pioneering study in Ref. \cite{Ishiguro}, which reported the 
formation of a double layer \cite{DL1,DL2,DL3} that cannot form if the plasma 
expands into a vacuum. Our simulation examines also the dynamics of protons 
as a first step towards a simulation of the mix of oxygen and nitrogen ions 
that constitute the residual gas in the physical experiment. Notable 
differences between the expansion of the hot and dense plasma into the 
ambient plasma and the expansion into a vacuum are observed.
 
The structure of this paper is the following. We describe the PIC method 
in Section 2 and we give the initial conditions and the simulation 
parameters. Section 3 models the initial phase of the plasma expansion 
at a high phase space resolution, revealing details of the electron 
expansion and of the quasi-equilibrium, which is established for the 
electrons. A double layer develops at the thermal pressure jump, which 
drags the electrons from the tenuous plasma into the hot plasma in form of 
a cool beam. The electrons from the hot plasma leak into the cool plasma, 
which reduces the temperature difference between both plasmas. Section 4 
examines the proton dynamics. The ambient plasma modifies the proton 
expansion. The thermal pressure 
jump evolves into a shock, which moves approximately with the proton thermal 
speed of the hot plasma. If the plasma expands into a vacuum, then a plasma 
density change can only be accomplished by ion beams \cite{Mora2}, while the 
plasma is here compressed by the shock. The fastest protons in our simulation 
form a beam that outruns the shock. It interacts with the protons of the 
ambient medium to form phase space holes in the electron and proton 
distributions. The proton phase space holes develop into a secondary shock 
ahead of the primary one. This process may result in secondary shocks in 
experiments, similar to the radiation-driven ones \cite{SecondShock}. The 
results are summarized in Section 5.

\section{The PIC simulation method and the initial conditions}

A PIC code approximates a plasma by an ensemble of computational particles 
(CPs), each of which is representing a phase space volume element. Each CP 
follows a phase space trajectory that is determined through the Lorentz 
force equation by the electric field $\mathbf{E}(\mathbf{x},t)$ and the 
magnetic field $\mathbf{B}(\mathbf{x},t)$. Both fields are evolved 
self-consistently in time using the Maxwell's equations and the macroscopic 
current $\mathbf{J}(\mathbf{x},t)$, which is the sum over the microcurrents 
of all CPs. The standard PIC method considers only collective interactions 
between particles, although some collisional effects are introduced through 
the interaction of CPs with the field fluctuations \cite{Dupree}. 

Collision operators have been prescribed for PIC simulations 
\cite{Jones,Sentoku}. The structures in the addressed experiment form and 
evolve in a plasma, in which collisional effects are not strong and such 
operators are thus not introduced here. We may illustrate this with the 
help of the electron collision rate $\nu_e \approx 2.9 \times 10^{-6} \, 
n_e \, \ln \Lambda \, T_e^{-3/2} \, \rm{s}^{-1}$ and the ion collision rate 
$\nu_i \approx 4.8 \times 10^{-8} Z^4 \mu^{-1/2} \, n_i \, \ln \Lambda \, 
T_i^{-3/2} \, \rm{s}^{-1}$ \cite{NRL} for a spatially uniform plasma with the 
number density $n_e = n_i = 10^{15}$ ($\rm{cm}^{-3}$) and the temperature 
$T_e = T_i = 10^3$ (eV). We take a Coulomb logarithm $\ln \Lambda = 10$ and 
we consider oxygen with $\mu = 16$. Both collision rates are comparable, if 
the mean ion charge $Z \approx 4$. We assume $\nu_e \approx \nu_i$. The 
electron plasma frequency $\omega_p \approx 10^{12} \, \rm{s}^{-1}$ gives 
the low relative collision frequency $\nu_e / \omega_p \approx 10^{-6}$. The 
plasma flow in the experiment and other aspects, which are not taken into 
account by this simplistic estimate, alter this collision frequency. The 
mean-free path has been estimated to be of the order of a cm \cite{Romagnani} 
and the ion beam with the speed $4 \times 10^5$ m/s crosses this distance 
during the time $\omega_p t \approx 25000$. This presumably forms the upper 
time limit, for which we can neglect collisions.   
 
The presence of particles with keV energies and the preferential expansion 
direction of the plasma in the experiment imply, that multi-dimensional PIC 
simulations should be electromagnetic in order to resolve the potentially 
important magnetic Weibel instabilities, which are driven by thermal 
anisotropies \cite{Weibel}. Such instabilities can grow in the absence of 
relativistic beams of charged particles, but they are typically weaker than 
the beam-driven ones \cite{Stockem}. Here we restrict our simulation to one 
spatial dimension $x$ (1D) and we set $\mathbf{B} (x,t=0)$. The plasma expands 
along $x$ and all particle beams will have velocity vectors aligned with $x$. 
The magnetic beam-driven instabilities have wavevectors that are oriented 
obliquely or perpendicular to the beam velocity vector and they are not 
resolved by a 1D simulation. The wavevectors, which are destabilized by the 
Weibel instability, can be aligned with the simulation direction, but only 
if the plasma is cooler along $x$ than orthogonally to it. Such a thermal 
anisotropy can probably not form. Our electromagnetic simulation confirms 
that no magnetic instability grows. The ratio of the magnetic to the total 
energy remains at noise levels below $10^{-4}$.

A 1D PIC simulation should provide a reasonable approximation to those
sections of the expanding plasma front observed in Ref. \cite{Romagnani}, 
which are planar over a sufficiently wide spatial interval orthogonal to 
the expansion direction. We set the length of the 1D simulation box to $L$. 
The plasma 1 is consisting of electrons (species 1) and protons (species 2), 
each with the density $n_h$ and the temperature $T_h$ = 1 keV, and it fills 
up the half-space $-L/2 < x < 0$. A number density $n_h = 10^{15} \,
\rm{cm}^{-3}$ should be appropriate with regard to the experiment. The 
half-space $0 < x < L/2$ is occupied by the plasma 2, which is composed of 
electrons (species 3) and protons (species 4) with the temperature $T_c$ = 
10 eV and the density $n_c = n_h / 100$. All plasma species have initially 
a Maxwellian velocity distribution, which is at rest in the simulation frame. 

The ablated target material drives the plasma expansion but its ions are 
probably not involved in the evolution of the shock and of the other plasma 
structures. These structures are observed already 100-200 ps after the laser 
impact at a distance of about 1 mm from the target. Aluminium ions, which 
are with a mass $m_A$ the lightest constituents of the target material, 
would have the thermal speed ${(T/m_A)}^{1/2} \approx 10^5$ m/s for $T$ = 
1 keV. Hundred times this speed or a temperature of 10 MeV would be 
necessary for them to propagate 1 mm in 0.1 ns. We thus assume here that 
the shock and the other plasma structures involve only the ions of the 
residual gas, which is air at a low pressure. If we assume that these ions 
have a high ionization state and comparable charge-to-mass ratios, then the 
protons may provide a reasonable approximation to their dynamics.

The equations solved by the PIC code are normalized with the number density 
$n_h$, the plasma frequency $\Omega_{1} = {(n_h e^2 / m_e \epsilon_0)}^{1/2}$ 
and the Debye length $\lambda_D = v_{t1} / \Omega_{1}$ of species 1, which 
equals that of the other species. The thermal speeds of the respective 
species are $v_{tj} = {(T_j / m_j)}^{1/2}$, where $j$ is the species 
index. We express the charge $q_k$ and mass $m_k$ of the $k^{th}$ CP in 
units of the elementary charge $e$ and electron mass $m_e$. Quantities 
in physical units have the subscript $p$ and we substitute $\mathbf{E}_p = 
\Omega_{1} v_{t1} m_e \mathbf{E} / e$, $\mathbf{B}_p = \Omega_{1} m_e \mathbf{B}
/ e$, $\mathbf{J_p} =e v_{t1} n_h \mathbf{J}$, $\rho_p = e n_h \rho$, $x_p =
\lambda_D x$, $t_p = t / \Omega_1$, $\mathbf{v}_p = \mathbf{v}v_{t1}$ and 
$\mathbf{p}_p = m_e m_k v_{t1} \mathbf{p}$. The 1D PIC code solves with 
$\tilde{v}_{t1} = v_{t1}/c$ the equations
\begin{eqnarray}
\nabla \times \mathbf{B} = \tilde{v}^2_{t1} \left (\partial_t \mathbf{E} +
\mathbf{J} \right ) , \, \, \nabla \times \mathbf{E}
= -\partial_t \mathbf{B}, \, \,
\nabla \cdot \mathbf{E} = \rho , \, \, \nabla
\cdot \mathbf{B} = 0, \\
\partial_t \mathbf{p}_{k} = q_{k} \left( \mathbf{E}[x_k] + 
\mathbf{v}_k \times \mathbf{B}[x_k] \right ), \, \, \rmd_t x_k = v_{k,x}.
\end{eqnarray}
The Lorentz force is solved for each CP with index $k$, position $x_k$ and 
velocity $\mathbf{v}_k$. It is necessary to interpolate the electromagnetic 
fields from the grid to the particle position to update $\mathbf{p}_k$ and 
the microcurrents of each CP have to be interpolated to the grid to update 
the electromagnetic fields. Interpolation schemes are detailed in Ref. 
\cite{Dawson}. Our code is based on the virtual particle electromagnetic 
particle-mesh method \cite{Eastwood} and it uses the lowest possible
interpolation order possible with this scheme. Our code is parallelized
through the distribution of the CPs over all processors.
  
Simulation 1 (Section 3) resolves the box length $L_S=3350$ by $N_S = 5 
\times 10^3$ grid cells of size $\Delta_{xS} = 0.67 \lambda_D$. The dense 
species 1 and 2 are each resolved by $8 \times 10^4$ CPs per cell and the 
tenuous species 3 and 4 by 800 CPs per cell, respectively. The simulation 
is evolved in time for the duration $t_S = 800$, subdivided into 45000 time 
steps $\Delta t_S$. Simulation 2 in Section 4 resolves the box length $L_L 
= 10 \, L_S$ by $N_L = 2.5 \times 10^4$ grid cells of size $\Delta_{xL} = 
1.34 \lambda_D$. This grid cell size is sufficiently small to avoid a 
significant numerical self-heating \cite{Birdsall} of the plasma during the 
simulation time. The total energy in the simulation is preserved to within 
$\approx 10^{-5}$. The species 1 and 2 are approximated by 6400 CPs per cell 
each and the species 3,4 by 64 CPs per cell, respectively. The system is 
evolved during $t_L = 25500$ with $6.4 \times 10^5$ time steps. 

We use periodic boundary conditions for the particles and the fields in all 
directions. Ideally, no particles or waves should traverse the full box 
length during the simulation duration. The group velocity for the
electrostatic waves and the propagation speed of the electrons are both
comparable to $v_{t1}$. We obtain $v_{t1}t_S / L_S \approx 0.24$ for 
simulation 1 and $v_{t1} t_L / L_L \approx 0.76$ for simulation 2.
Both simulations ran on 16 CPUs on an AMD Opteron cluster (2.2 GHz). 
Simulation 1/2 ran for 100/800 hours.

\section{Simulation 1: Initial development}

Our initial conditions involve a jump in the bulk plasma properties at 
$x\approx 0$. Some electrons of the plasma 1 will expand into the 
half-space $x>0$ occupied by the plasma 2. The slow protons can not
keep up with the electrons and the resulting charge imbalance gives rise 
to an electrostatic field $E_x$. This $E_x$ confines the electrons of 
plasma 1 and it accelerates the electrons from the plasma 2 into the 
half-space $x<0$. The electrons of the plasma 1 and 2 with $x<0$ are 
separated along the velocity direction by the electrostatic potential
and form a double layer. 

\begin{figure}
\begin{center}
\includegraphics[width=7.5cm]{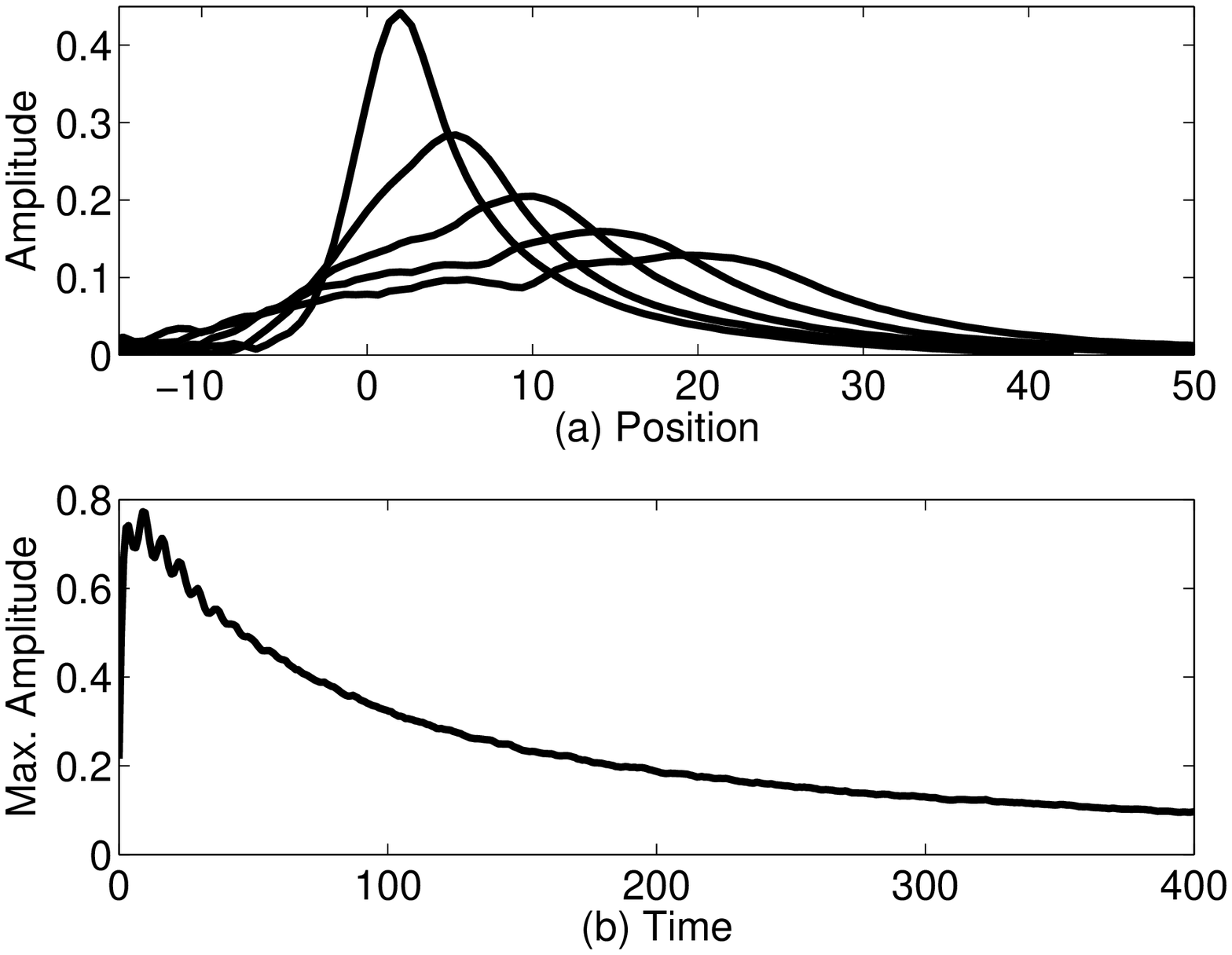}
\includegraphics[width=7.5cm]{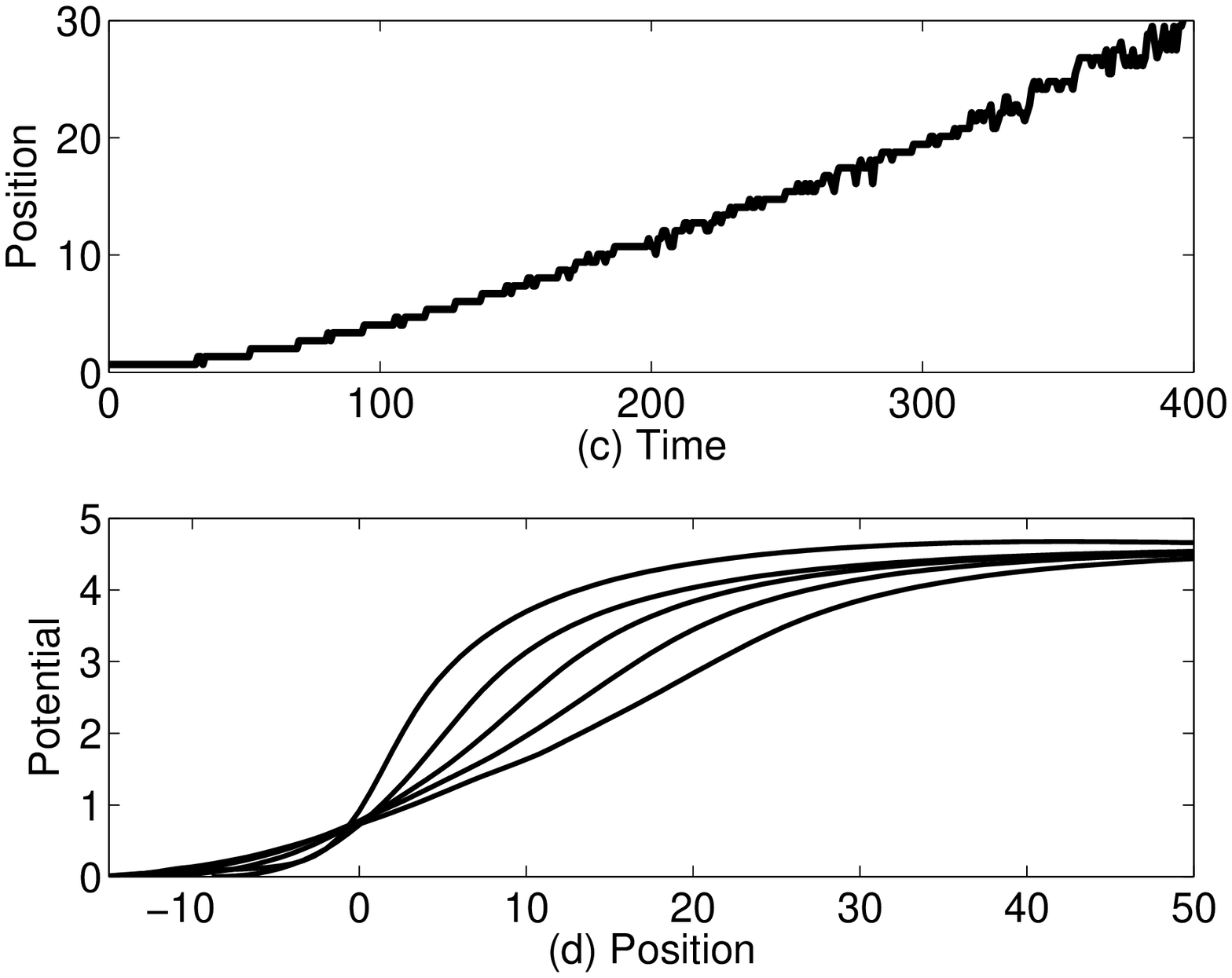}
\caption{The electric field: (a) shows $E_x$ at the times $t=60,120,180,240$
and $300$. The maximum amplitude decreases with time as (b) is showing and 
the location of the electric field maximum moves towards positive $x$ (c). 
The potential in kV obtained from the $E_x$ distributions from (a) is displayed 
in (d). The potential jump remains unchanged, but the gradient is eroded 
with the time.\label{fig1}}
\end{center}
\end{figure}

Figure \ref{fig1} examines the $E_x$ and its potential.
The amplitude of $E_x$ peaks initially at $x \approx 0$ and it accelerates 
the electrons into the negative $x$-direction. The position of the maximum 
of $E_x$ moves to larger $x$ with increasing times and the peak amplitude
decreases. The spatial profile of $E_x$ is smooth, which contrasts the one
that drives the plasma expansion into a vacuum that has a cusp 
\cite{Mora2}. The potential difference $\approx$ 5 kV between plasma 1 and 2 
remains unchanged. The spatial interval, in which the amplitude of $E_x$ is 
well above noise levels, is bounded. An interesting property of the double 
layer can thus be inferred according to \cite{DL3}. Its electrostatic field 
can only redistribute the momentum between the four plasma species, but it 
can not provide a net flow momentum. This is true if the double layer is 
one-dimensional and electrostatic. The decrease of the peak electric field 
in Fig. \ref{fig1}(b) resembles that in Fig. 3 in Ref. \cite{Mora}. The 
decreasing electric force, in turn, implies that the ion acceleration in 
the Fig. 4 of Ref. \cite{Mora} decreases as the time progresses, which
should hold for our simulation too. 

\begin{figure}
\begin{center}
\includegraphics[width=7.5cm]{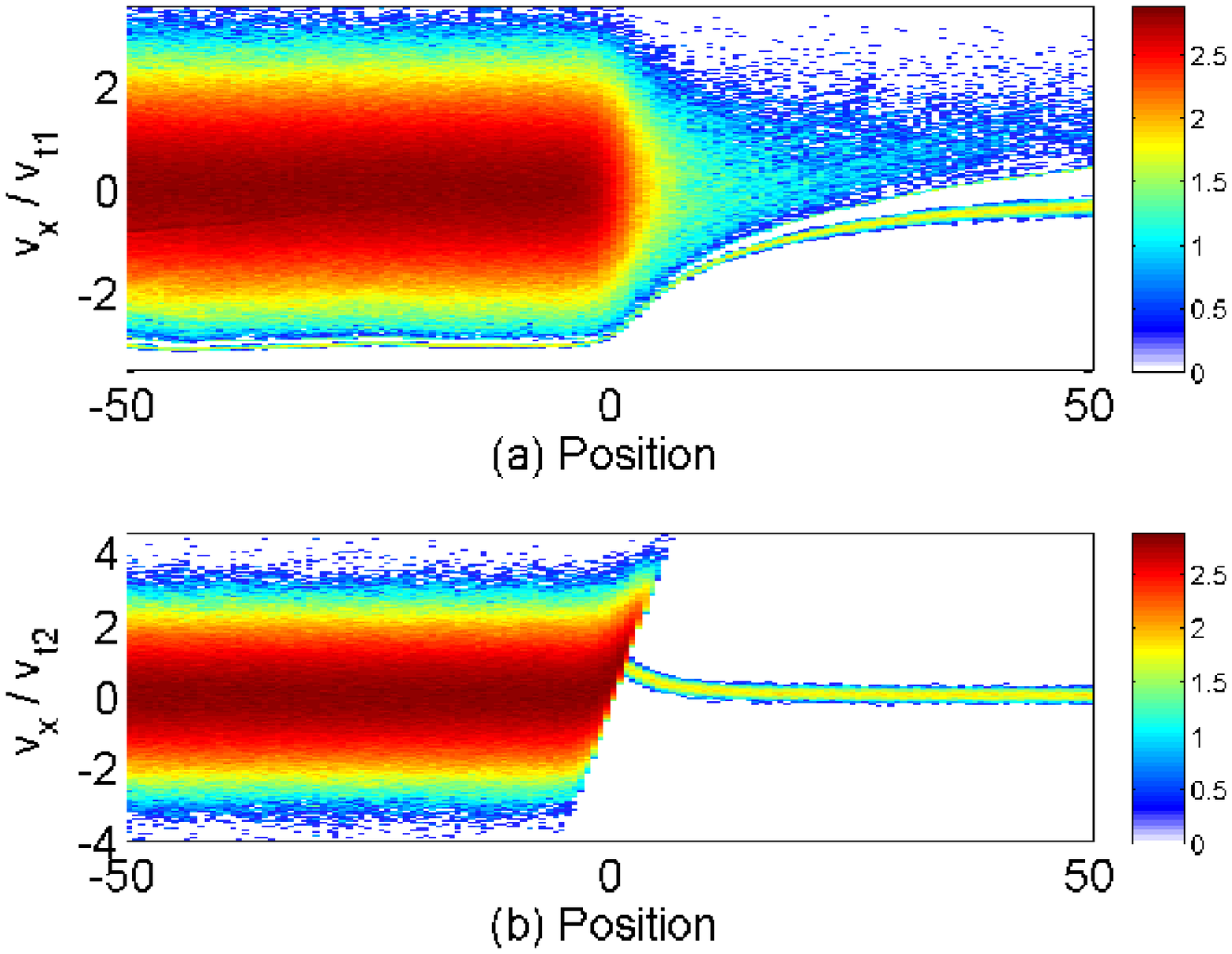}
\includegraphics[width=7.5cm]{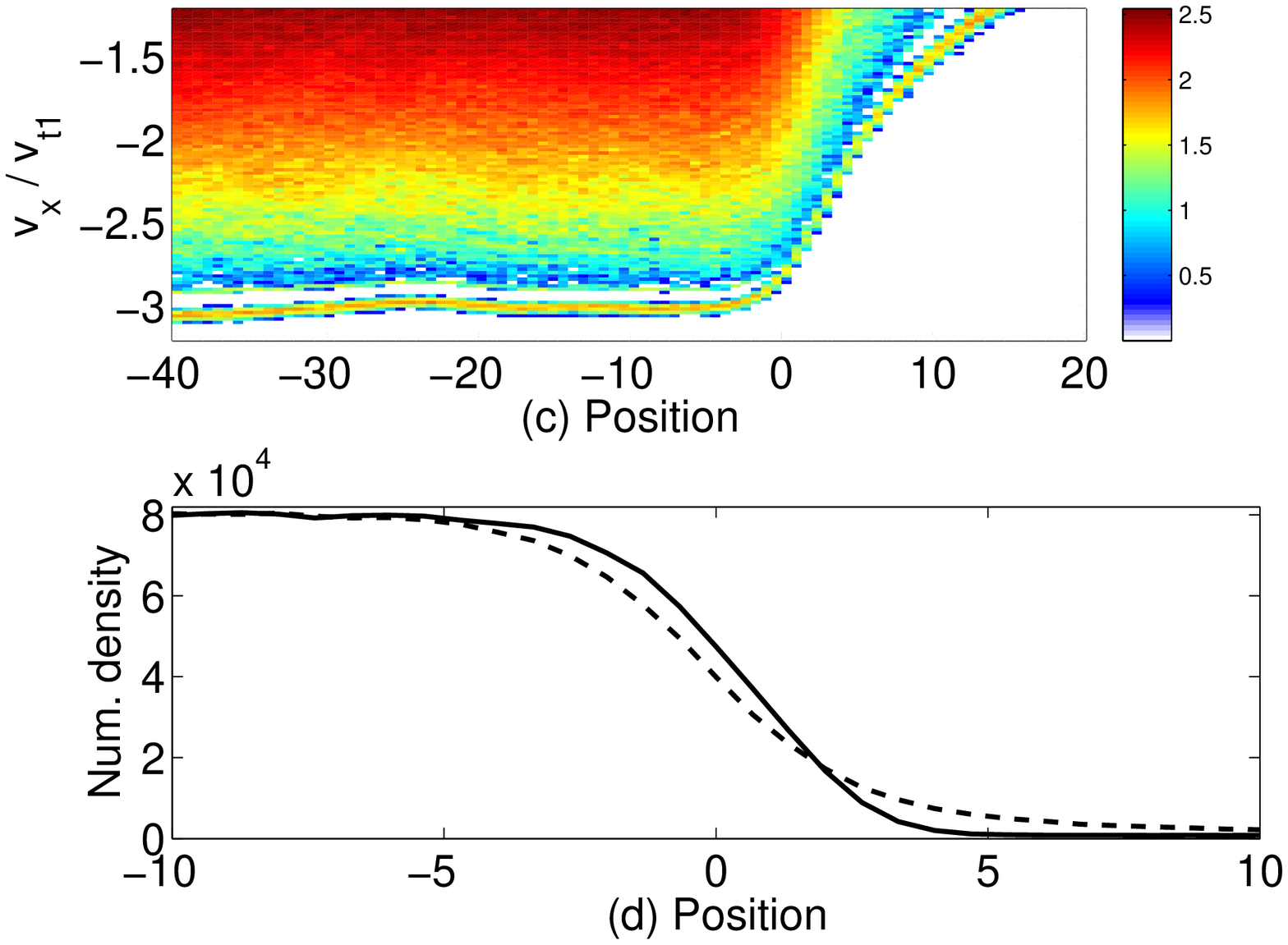}
\caption{(Colour online) The plasma distribution at $t=60$: (a) shows the 
electron phase space distribution. Most electrons from the dense plasma 
remain confined to $x<0$, but some diffuse into the tenuous plasma. (b) 
shows the proton phase space distribution. Some protons with $v_x > 0$ 
are accelerated in $0<x<5$. The protons with $x,v_x< 0$ stream freely to 
lower values of $x$. (c) The electron phase space distribution 
reveals a double layer. (a-c) show the 10-logarithmic number of CPs. 
(d) shows the number of CPs per cell of the electrons (dashed curve) and 
the protons (solid curve).\label{fig2}}
\end{center}
\end{figure}

The plasma phase space distribution at $t=60$ is investigated in Fig.
\ref{fig2}.
A tenuous hot beam of electrons is diffusing from the plasma 1 into the 
half-space $x>0$, while the mean speed of the electrons of the plasma 2 
becomes negative. The electrons of plasma 1 and 2 with $x<0$ are separated 
by a velocity gap of $\approx v_{t1}/10$. The protons that were close to 
the initial boundary $x=0$ at $t=0$ have propagated until $t=60$ for a 
distance, which is proportional to their speed. A sheared velocity 
distribution can thus be seen in Fig. \ref{fig2}(b). The fastest protons 
of the plasma 1 with $x>0$ have also been accelerated by the $E_x$ by 
about $v_{t2}/2$, reaching now a peak speed $\approx 4v_{t2}$. The fastest
protons are found to the right of the maximum of $E_x$ at $x\approx 2$ at
$t=60$ in Fig.\ref{fig1}(a). A similar acceleration is observed for the 
protons of plasma 2 in $0<x<5$. The densities of the electrons and protons 
disagree in the interval $-5<x<5$ and the net charge results in the 
electrostatic field $E_x > 0$. Both curves in Fig. \ref{fig2}(d) intersect 
at $x\approx 2$, which coincides with the position in Fig. \ref{fig1}(a), 
where the $E_x$ has its maximum at $t=60$. 

The density of the cold protons in Ref. \cite{Mora2} is practically 
discontinuous at the front of the expanding plasma, while it changes 
smoothly in our simulation. This is a result of our high proton temperature, 
which causes the thermal diffusion of the protons. The contour lines of the 
electron phase space density are curved at $x\approx 0$. Most electrons of
plasma 1 that move to increasing values of $x$ are reflected by the 
electrostatic potential at $x\approx 0$. These density contour lines resemble 
those of the distribution of electrons that expand into a vacuum at an early 
time in Ref. \cite{Mora2}, which are all reflected by the potential at the
plasma front. Here the inflow of electrons from plasma 2 into plasma 1 allows
some of the electrons of plasma 1 to overcome the potential. The electrons
provide all energy for the proton expansion in Ref. \cite{Mora2} and their
distribution develops a flat top. Here the proton thermal energy is the main 
driver and consequently the electron velocity distribution shows no clear 
deviation from a Maxwellian at any time.

\begin{figure}
\begin{center}
\includegraphics[width=7.5cm]{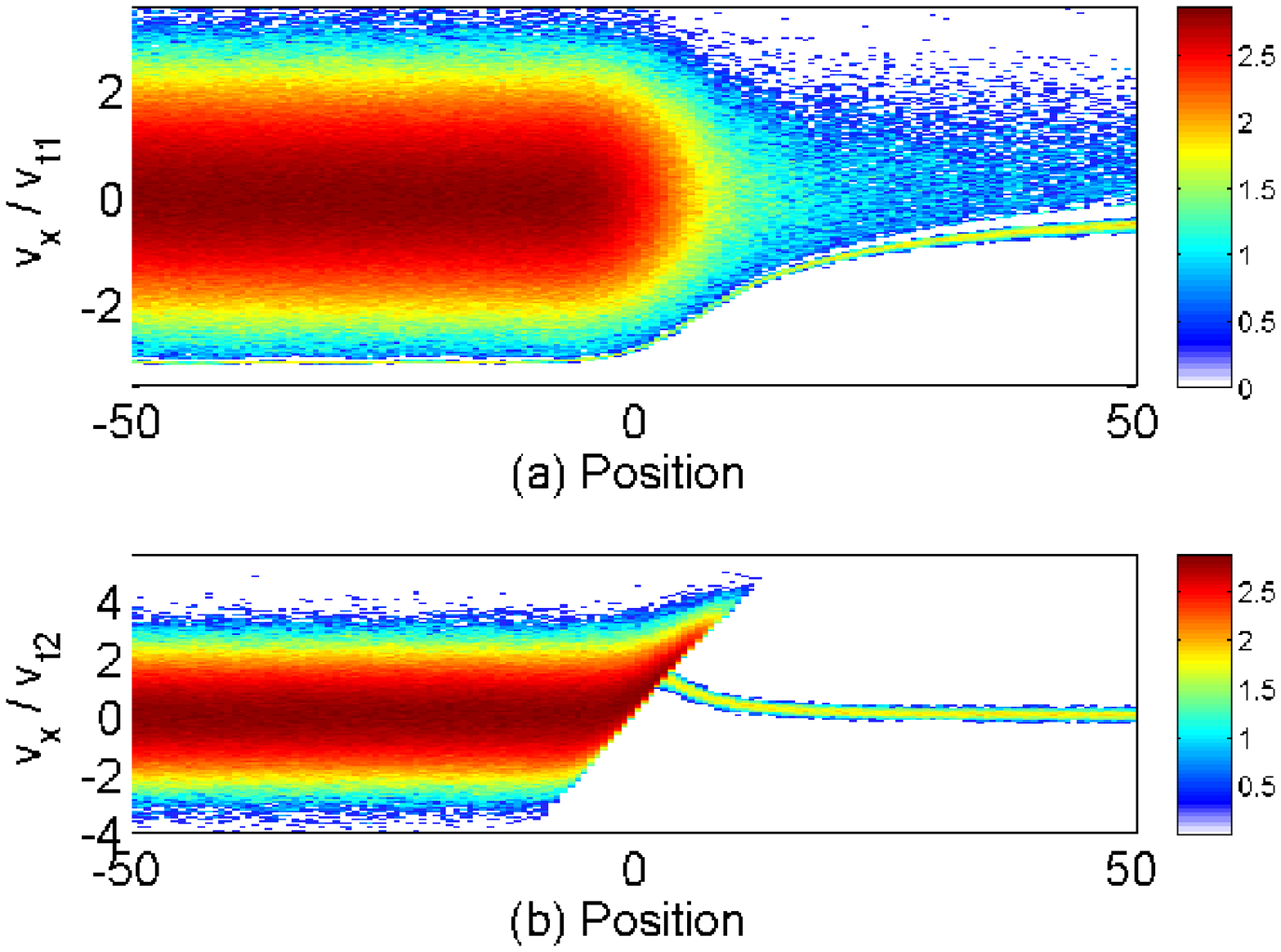}
\includegraphics[width=7.5cm]{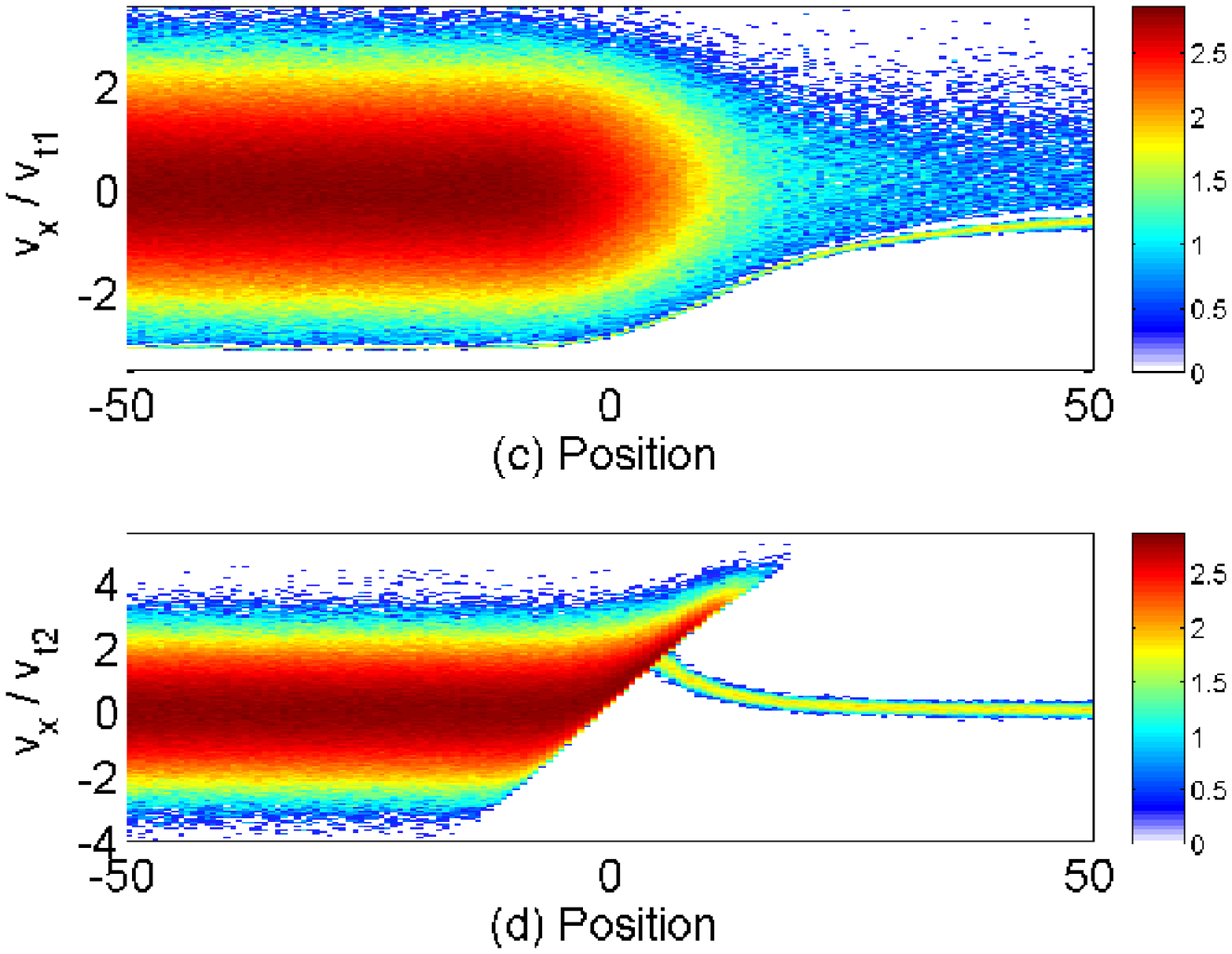}
\caption{(Colour online) The 10-logarithmic phase space densities in units 
of CPs: The electron distribution in (a) and the proton distribution in (b) 
are sampled at $t=120$, while (c) and (d) show them at $t=180$.
The protons in the interval $x,v_x < 0$ convect almost freely away from 
$x=0$. The protons of the dense plasma in $x,v_x > 0$ accelerate. Electrons 
diffuse from the plasma 1 into the plasma 2 and form a hot beam, while 
electrons from the plasma 2 enter the plasma 1 in form of a cold beam.
\label{fig3}}
\end{center}
\end{figure}

Figure \ref{fig3} shows the plasma phase space distributions at the times 
$t=120$ and $t=180$. 
The plasma distributions are qualitatively similar to that in Fig. \ref{fig2}. 
Electrons diffuse out from plasma 1 into plasma 2, forming a hot beam, while 
the electrons of plasma 2 are dragged into the half-space $x<0$ in form of a 
cold beam. The confined electrons of plasma 1 expand to increasing $x$ at a 
speed, which is determined mainly by the protons. The proton distribution 
shows an increasing velocity shear, but the apparent phase space boundary 
between the protons of plasma 1 and 2 is still intersecting $v_x =0$ at $x=0$.
The front of the protons of plasma 1 at $t=120$ and $t=180$ is close to 
the position of the maximum of $E_x$ in Fig. \ref{fig1}(a) at $x\approx 5$ 
for $t=120$ and $x\approx 10$ for $t=180$. The protons at the front of 
plasma 1 and the protons of plasma 2 in the same interval are accelerated 
by the $E_x > 0$ and reach the peak speed $\approx 5 v_{t2}$.

\begin{figure}
\begin{center}
\includegraphics[width=7.5cm]{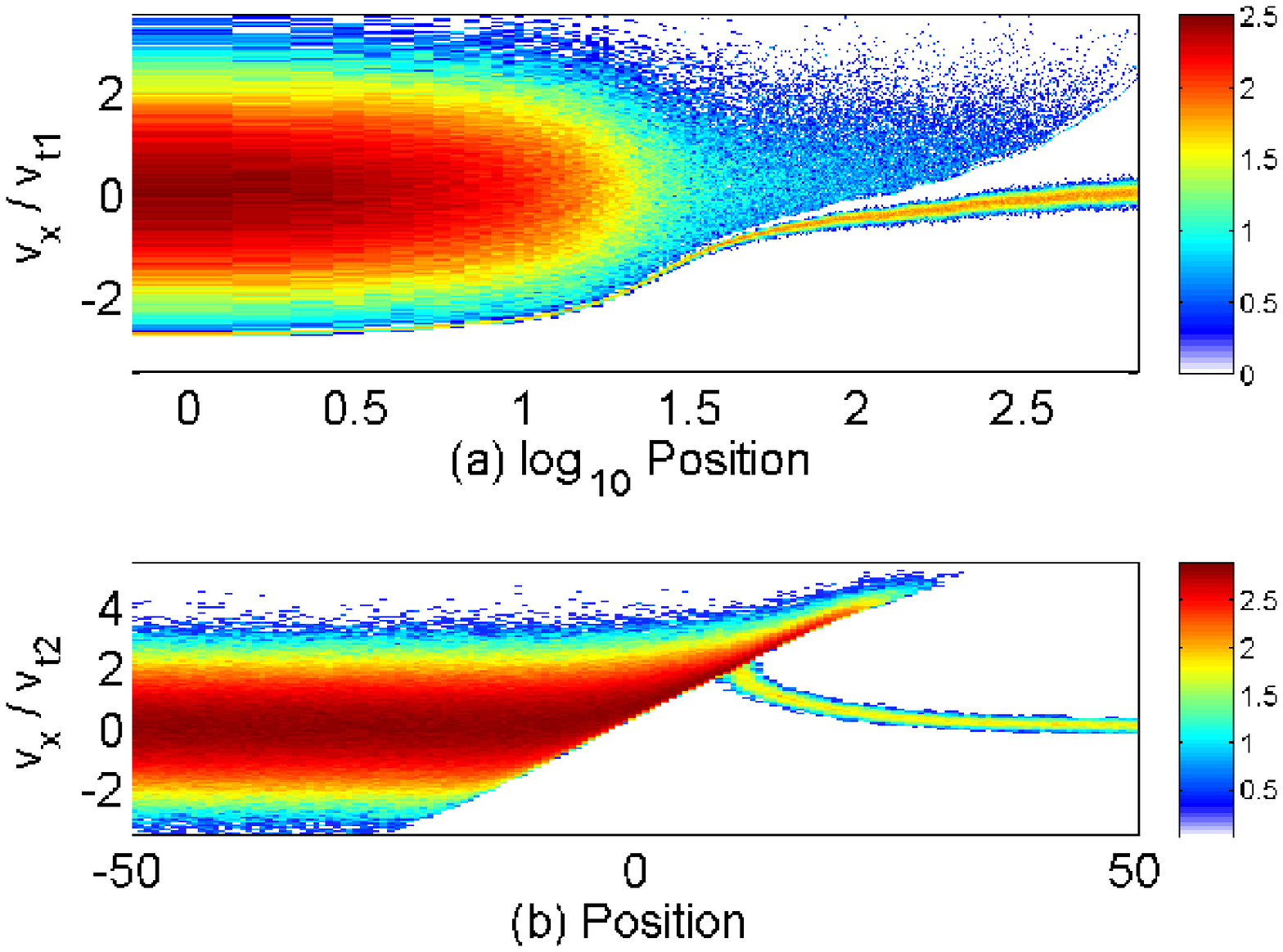}
\includegraphics[width=7.5cm]{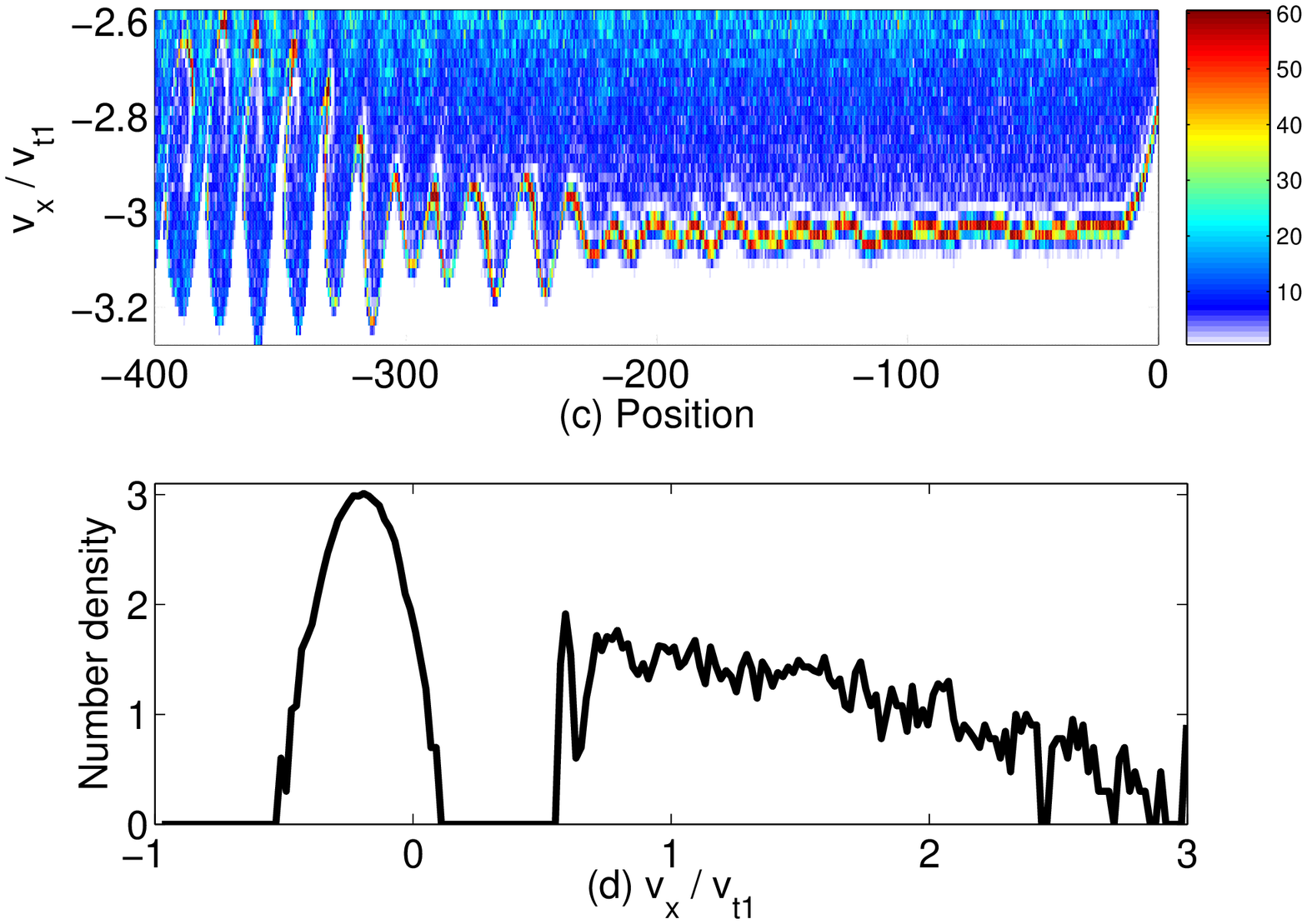}
\caption{(Colour online) The 10-logarithmic number of CPs representing the 
electrons (a) and the protons (b) at the time $t=300$. The electrons of plasma 
1 have spread out to $x\approx 700$. The protons of plasma 1 with $x>0$ are 
accelerated to about $5v_{t2}$. The electron density in units of CPs for $x<0$
is displayed in (c). Electron phase space holes are present for $x<-300$. The 
electron distribution integrated over $250<x<260$ is shown in (d).\label{fig4}}
\end{center}
\end{figure}

The electrons of plasma 1 in Fig. \ref{fig4} at $t=300$ have expanded into 
the half-space $x>0$ for several hundred Debye lengths. The electrons from 
the plasma 2, which have been dragged towards $x<0$, interact with the 
electrons of plasma 1 through a two-stream instability. A chain of large 
electron phase space holes has developed for $-400 < x < -300$, which 
thermalize the beam distribution. No two-stream instability is yet observed 
in the interval $x>0$, even though a beam distribution is present, for 
example, at $x\approx 250$. The change of the mean speed of the electron 
beam leaked from plasma 1 for $x>0$ inhibits the resonance that gives rise 
to the two-stream instability. The mean speed of the electrons of plasma 2 
does not vanish any more and it varies along $x>0$ to provide the return 
current that cancels that of the electrons of plasma 1. The $E_x$ has 
noticably accelerated the protons in the interval $10<x<30$, which still
show the sheared distribution in the interval $-25<x<25$. 

The evolution of the plasma is animated in the movies 1 (electrons) and 2 
(protons). The axis labels $v_{eh} = v_{t1}$ and $v_{ph} = v_{t2}$. The 
colour scale denotes the 10-logarithmic number of CPs. The movie 1 reveals 
that a thin band of electrons parallel to $v_x$ propagates away instantly 
from the plasma 1 and towards $x>0$. These electrons leave the plasma 1, 
before the $E_x$ has grown. The electrons diffusing into $x>0$ at later 
times, when the $E_x$ has developed, form a tenuous beam with a broad 
velocity spread. The electrons of plasma 1 can overcome the double layer
potential of $\approx$ 5 kV if their speed is $v\geq 3 v_{t1}$ prior to the 
encounter of its electrostatic field. The movie 1 furthermore illustrates 
the growth of the two-stream instability between the electron beam originating 
from the plasma 2 and the confined electrons of plasma 1 in $x<0$ and its 
saturation through the formation of electron phase space holes. The movie 2 
demonstrates, how the velocity shear of the protons develops and how the 
fastest protons of plasma 1 in $x>0$ are accelerated by $E_x$. Neither the 
Fig. \ref{fig4} nor the movie 2 reveal the formation of a shocked proton
distribution prior to the time $t_S$.

\begin{figure}
\begin{center}
\includegraphics[width=7.5cm]{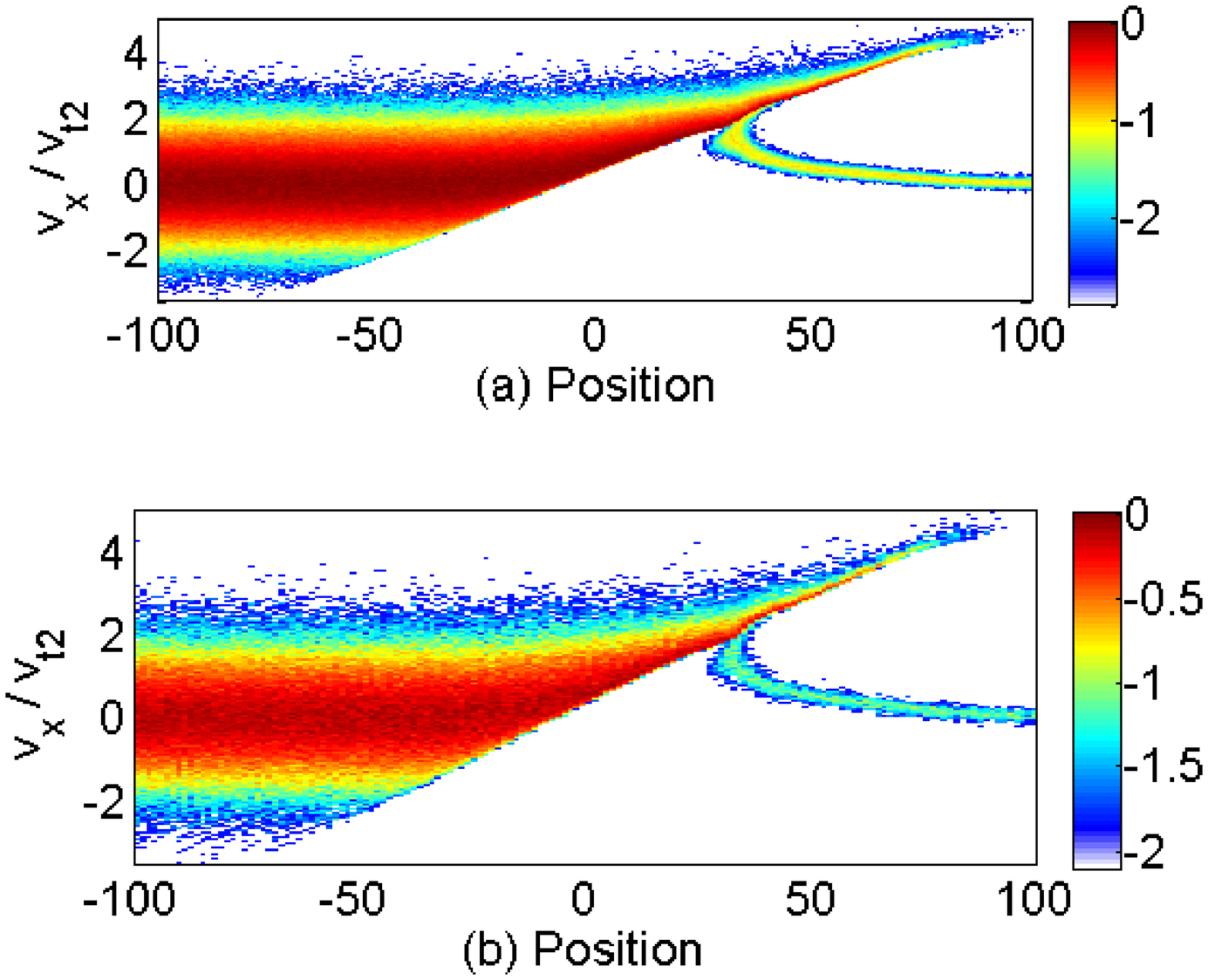}
\includegraphics[width=7.5cm]{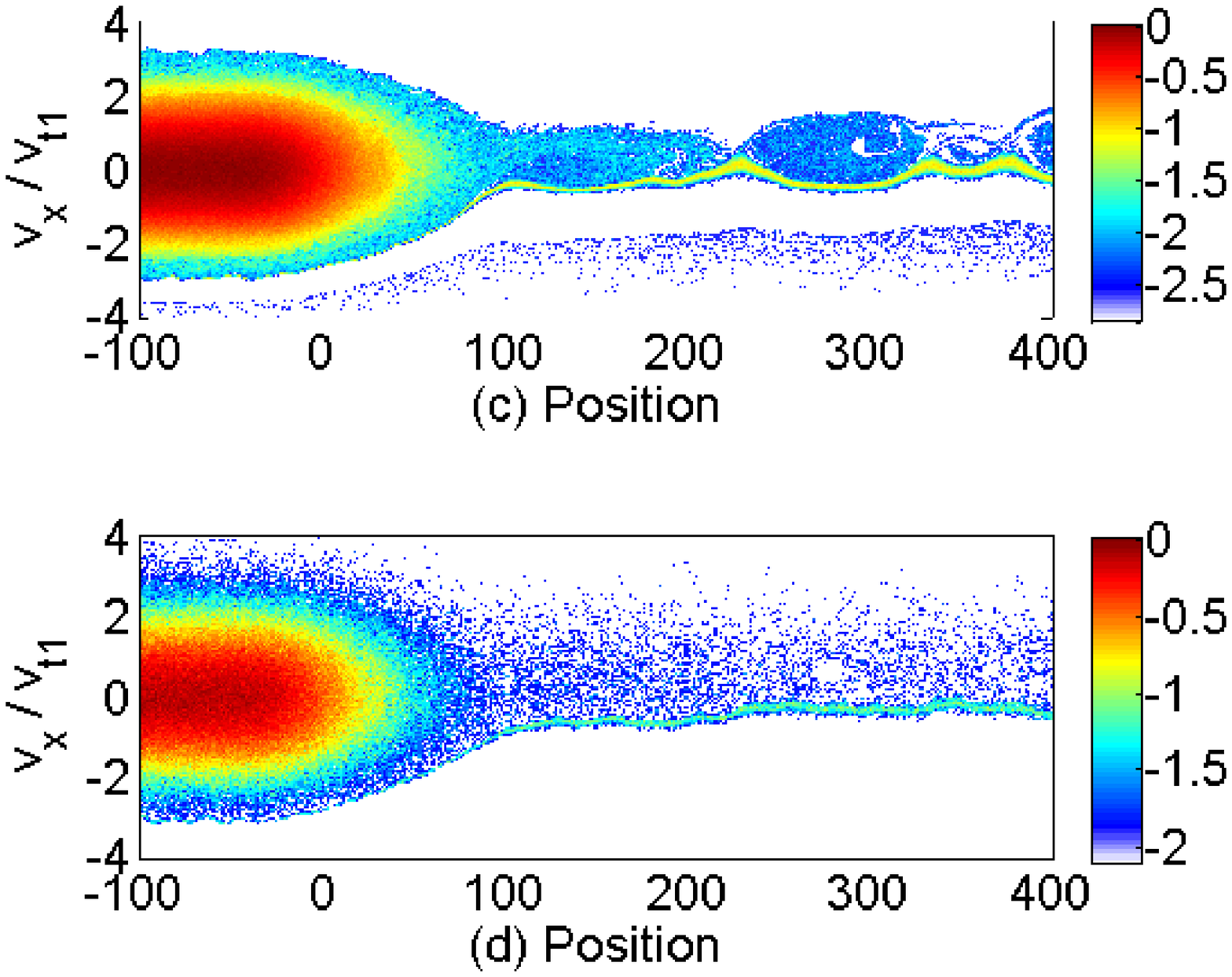}
\caption{(Colour online) The 10-logarithmic phase space distributions, 
normalized to their respective peak values: (a) shows the proton distribution 
in simulation 1 and (b) that in simulation 2. The electron distributions in 
the simulations 1 and 2 are displayed in (c) and (d), respectively.
\label{fig5}}
\end{center}
\end{figure}

We expand the simulation box and we reduce the statistical representation
of the plasma. Ideally, the plasma evolution should be unchanged. 
Figure \ref{fig5} compares the plasma 
data provided by simulation 1 (box length $L_S$) and by simulation 2 
($L_L = 10L_S$) at the time $t_S$, when we stop simulation 1. The proton 
distributions in both simulations are practically identical and we notice 
only one quantitative difference. The sheared proton distribution of 
plasma 1 extends to $x\approx -60$ and $v_x \approx -3 v_{t2}$ in simulation 1, 
while it reaches only $x\approx -50$ and $v_x \approx -2v_{t2}$ in simulation
2. This can be attributed to the better representation of the high-energy 
tail of the Maxwellian in simulation 1. 

The bulk electron distributions in both simulations agree well for $x<100$. 
The interaction of the confined electrons of plasma 1 with the expanding
protons is thus reproduced well by both simulations. We find a beam of 
electrons with $x>100$ and $v_x \approx -3v_{t1}$ in Fig. \ref{fig5}(c), 
which is accelerated by the double layer to $-4v_{t1}$ in the interval 
$-100 < x< 100$. This beam originates from the second boundary between the 
dense and the tenuous plasma at $x=L_S/2$ in simulation 1. It is thus 
an artifact of our periodic boundary conditions. Its density is three
orders of magnitude below the maximum one and it does thus not carry 
significant energy. This tenuous beam does not show any phase space 
structuring, which would be a consequence of instabilities, and it has
thus not interacted with the bulk plasma. Its only consequence is to provide
a weak current that should not modify the double layer. This fast beam is 
absent in Fig. \ref{fig5}(d), because the electrons could not cross
the distance $L_L/2$ in simulation 2 during the time $t_S$.

The electron distributions for $x>100$ and $v_x > 0$ computed by both 
simulations differ substantially. The electrons form phase space vortices 
in simulation 1, while the electrons in simulation 2 form a diffuse beam 
with some phase space structures, e.g. at $x\approx 300$. Phase space 
vortices are a consequence of an electrostatic two-stream instability, 
which must have developed between the leaked electrons of plasma 1 and the 
electrons of plasma 2. Only the electrons of plasma 1 with $v>3v_{t1}$ can 
overcome the double layer potential. These leaked electrons form a smooth 
beam in simulation 1 that can interact resonantly with the electrons of 
plasma 2 to form well-defined phase space vortices. The statistical 
representation of the leaking electrons in simulation 2 provides a minimum
density that exceeds the density of these vortices. 

\section{Simulation 2: Long term evolution}

\begin{figure}
\begin{center}
\includegraphics[width=7.5cm]{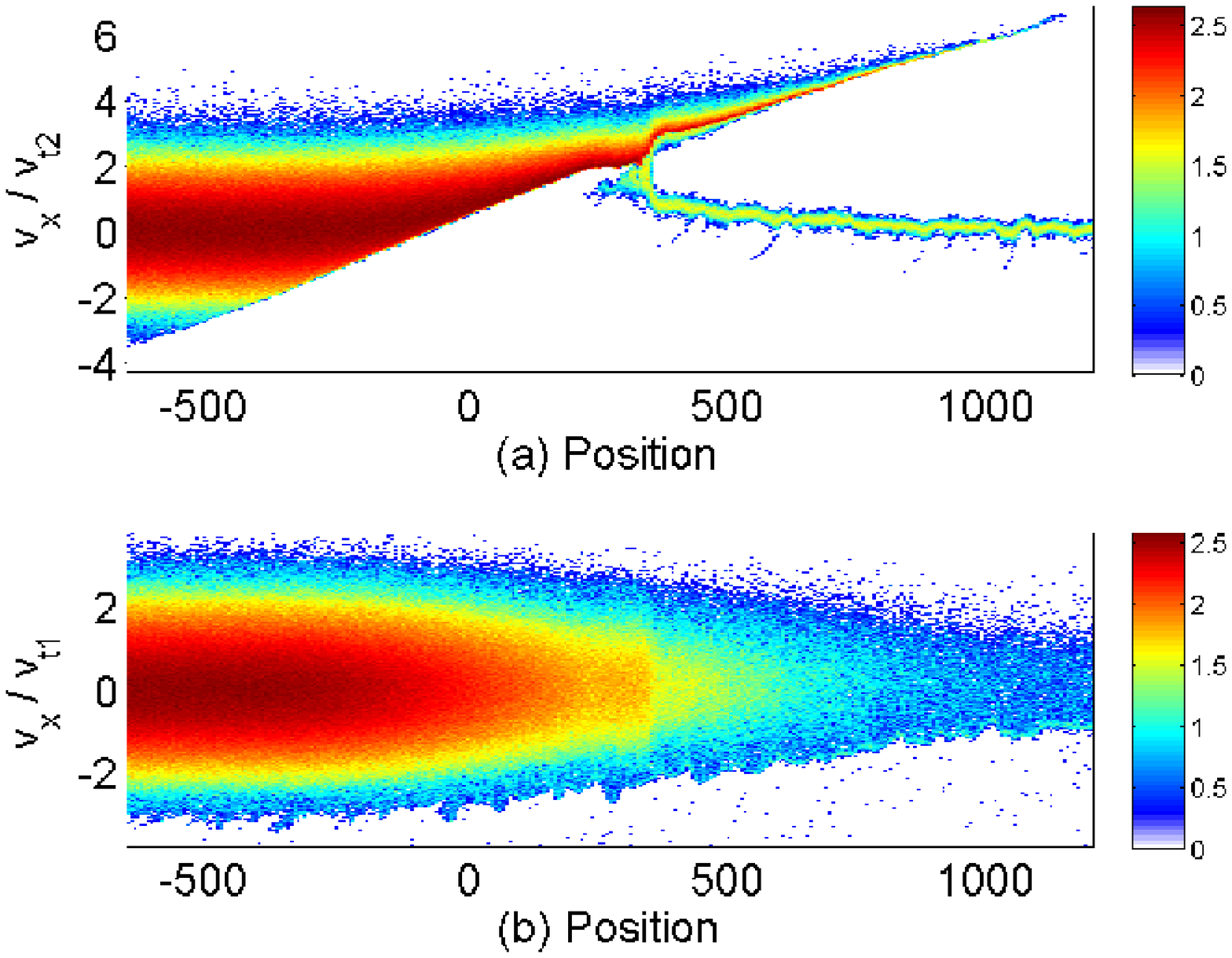}
\includegraphics[width=7.5cm]{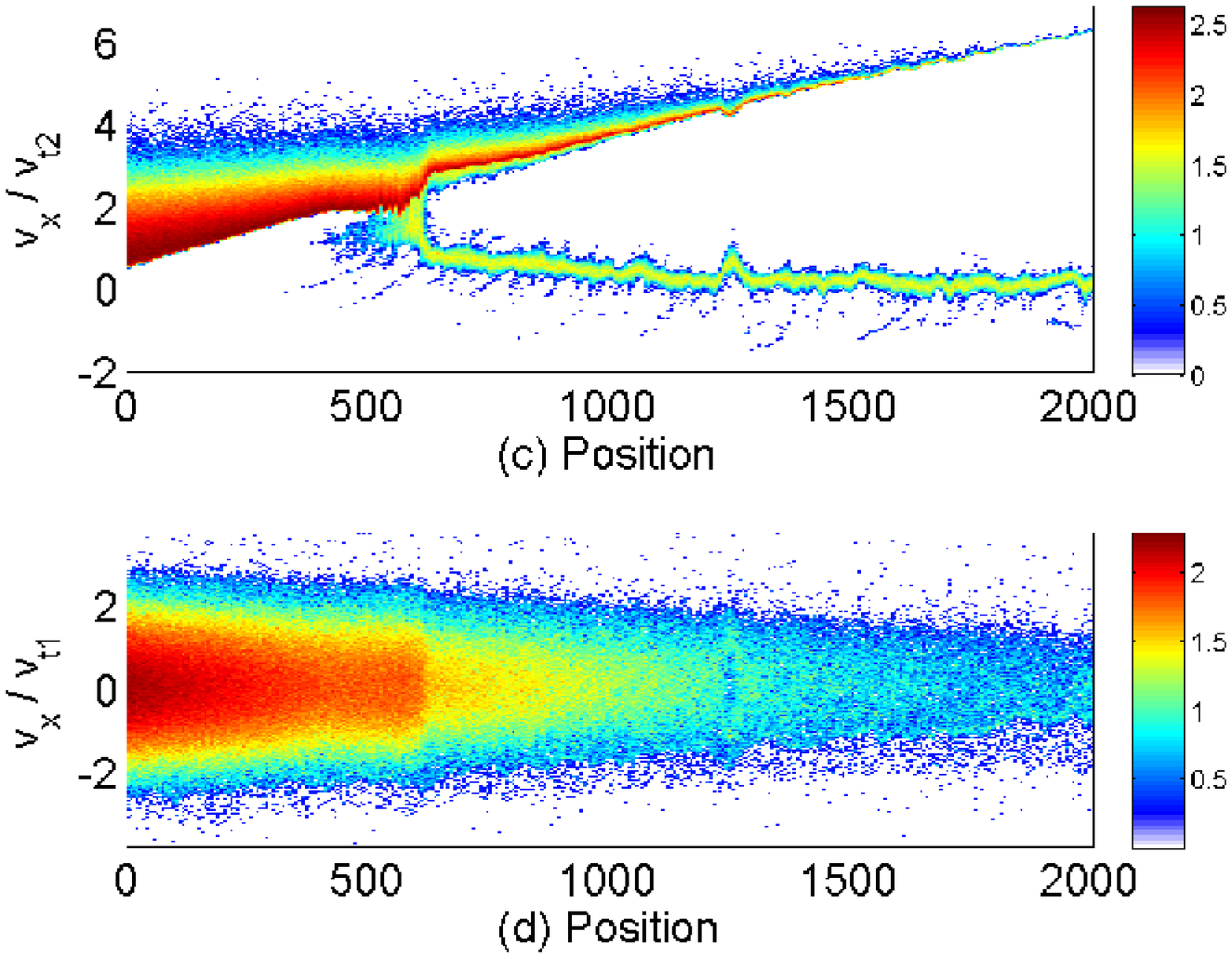}
\caption{(Colour online) The 10-logarithmic phase space distributions of the 
snapshots $S_1$ (a,b) and $S_2$ (c,d) in units of CPs: A shock is developing 
in the proton distribution (a) at $x\approx 300$. The electrons are distributed 
symmetrically around $v_x = 0$ in (b) and their density value jumps at 
$x\approx 300$. The proton shock in (c) and the electron density jump in (d) 
have propagated to $x \approx 600$ at the mean speed $\approx v_{t2}$.
\label{fig6}}
\end{center}
\end{figure}

We examine the plasma at three times. The snapshot $S_1$ corresponds to
the time $t = 8000$, $S_2$ to $t= 16000$ and $S_3$ to $t = 25500$. The plasma 
phase space distributions for $S_1$ and $S_2$ are displayed by Fig. \ref{fig6}.
The proton distribution is still qualitatively similar to that at $t=300$
in Fig. \ref{fig4}. The phase space boundary between the protons of plasma 1 
and 2 has been tilted further by the proton streaming. The key difference 
between the Figs. \ref{fig4} and \ref{fig6} is found, where the proton
distribution of the plasma 1 merges with that of the plasma 2. This 
collision boundary is located at $x\approx 300$ for $S_1$ and at $x\approx 
600$ for $S_2$, which evidences an approximately constant speed of this 
intersection point. The propagation speed is $\approx v_{t2}$. The protons 
directly behind this collision boundary, e.g. in $450<x<550$ for $S_2$, do 
not show a velocity shear. Their mean speed and velocity spread is spatially 
uniform in this interval, evidencing the downstream region of a shock. The 
upstream proton distribution with $x > 600$ for $S_2$ resembles, however, 
only qualitatively that of an electrostatic shock \cite{Forslund}. That
consists of the incoming plasma and the shock-reflected ion beam. The density 
of the beam with $v_x \approx 4 v_{t2}$ exceeds that of the plasma 2 in the 
same interval and its mean speed exceeds the $v_s \approx v_{t2}$ of the 
shock by the factor 4. A shock-reflected ion beam would move at twice 
the shock speed and its density would typically be less than that of the 
upstream plasma, which the shock reflects. The linear increase of the 
proton beam velocity with increasing $x$ is reminiscent of the plasma 
expansion into a vacuum \cite{Mora3}, but it is here a consequence of the 
shear introduced by the proton thermal spread.

The electron distribution at $t=t_S$ in Fig. \ref{fig5}(d) could be 
subdivided into the cool electrons of plasma 2 and the leaked hot electrons 
of plasma 1, while the electrons in the interval $x>750$ have a symmetric 
velocity distribution in Fig. \ref{fig6}(b) that does not permit such a 
distinction. The electron temperature gradient has also been eroded. The 
electron phase space density decreases by an order of magnitude as we go 
from $v_x = 0$ to $v_x \approx 2v_{t1}$ at $x\approx 0$ and at $x\approx 
2000$ in Fig. \ref{fig6}(d) and the thermal spread is thus comparable at
both locations. We attribute this temperature equilibration to electrostatic 
instabilities, which were driven by the electron beam that leaked through 
the boundary at $x=0$, and also to the electron scattering by the simulation 
noise. The noise amplitude is significant in the interval $x>0$ due to the 
comparatively low statistical representation of the plasma, in particular 
that of the hot leaked electrons.

The electron density jumps at both times in Fig. \ref{fig6} at the positions, 
where the protons of plasma 1 and 2 intersect. The electron distribution for 
$S_2$ furthermore shows a spatially uniform distribution in $450<x<550$, as 
the protons do. The electrons have thermalized and any remaining free energy 
would be negligible compared to that of the protons. The electron density 
merely follows that of the protons to conserve the plasma quasi-neutrality. 
This electron distribution thus differs from the similarly looking one, which 
has been computed at late times in Ref. \cite{Mora2}. There the electrons 
changed their velocity distribution in response to the energy lost to the 
protons.

\begin{figure}
\begin{center}
\includegraphics[width=7.5cm]{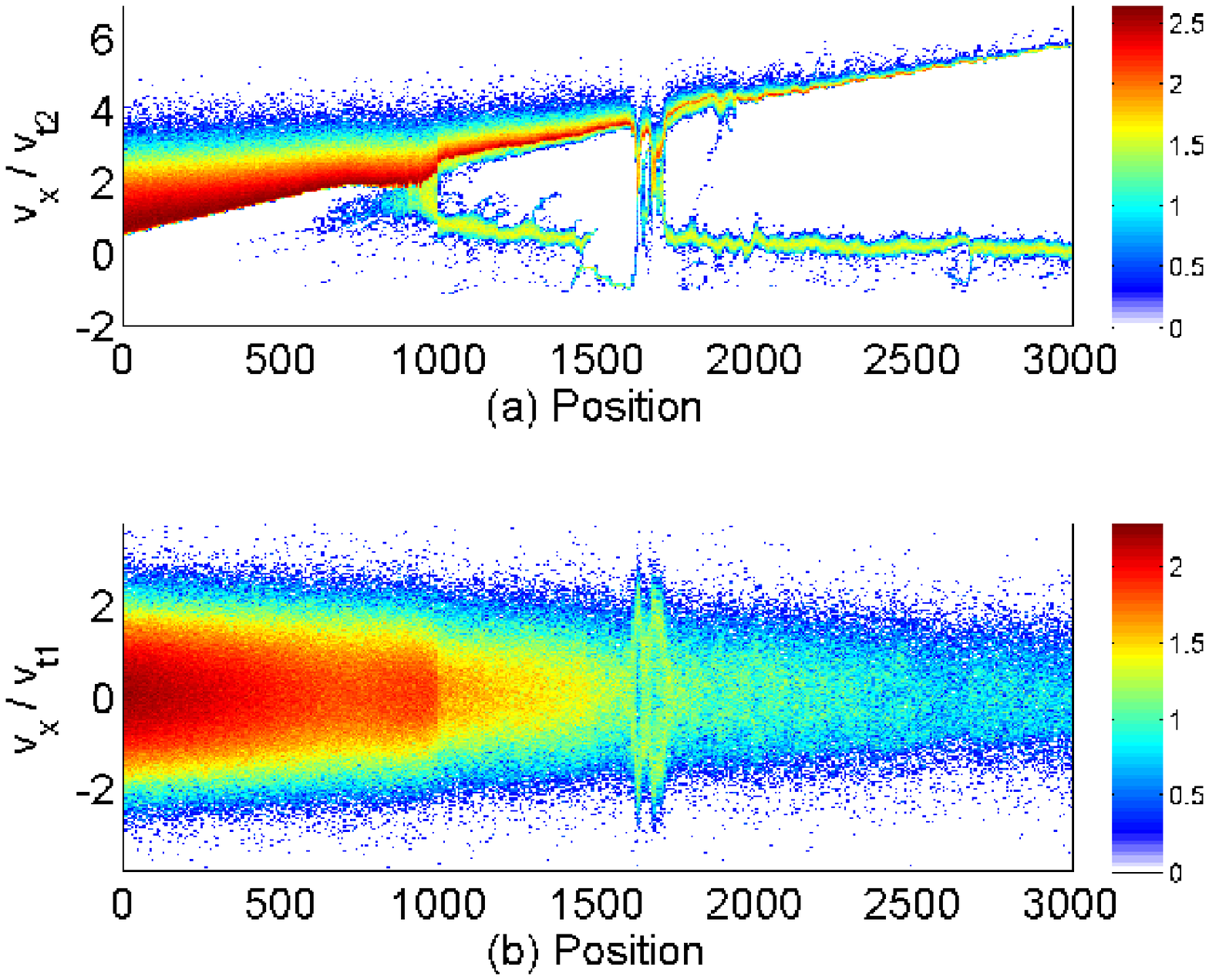}
\includegraphics[width=7.5cm]{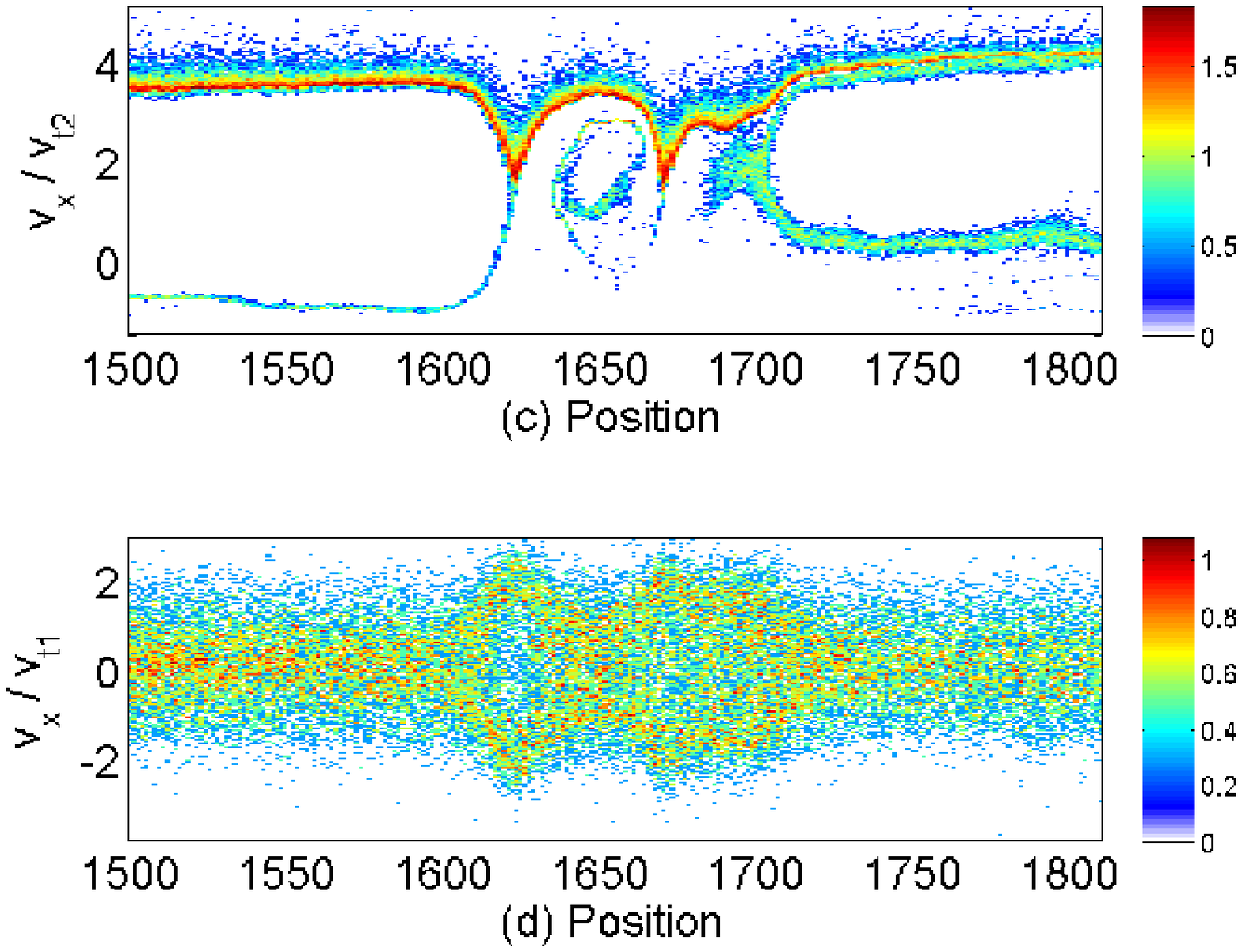}
\caption{(Colour online) The 10-logarithmic phase space density for 
$S_3$ in units of CPs: (a) displays the proton distribution and (b) the 
electron distribution. The shock is located at $x\approx 900$ and phase 
space holes develop in the proton (c) and electron (d) distribution at 
$1600 <x< 1700$. A new shock is growing at $x\approx 1700$ in (c).
\label{fig7}}
\end{center}
\end{figure}

The time $10 t_S$ corresponding to $S_1$ and the box length $L_L = 10L_S$ 
imply, that we should see some electrons emanated by the plasma boundary at 
$x=L_L/2$ as in Fig. \ref{fig5}. Only the electrons with $v<-2.1 v_{t1}$ 
would be fast enough to cross the interval $0<x<L_L/2$ occupied by plasma 2 
during the time $10t_S$. These electrons correspond to the few fast electrons 
in Fig. \ref{fig6}(b) with $x>0$ and $v<0$. An increased number of fast 
electrons moving in the negative $x$-direction is visible at the snapshot 
$S_2$. The electrons emanated from the plasma boundary at $x=L_L/2$ reach 
now the boundary at $x=0$ in significant numbers. The diffuse phase space 
distribution of these electrons implies, however, that they do not carry 
with them enough free energy that could result in instabilities that drive
strong electrostatic fields.

The shock structure and the density jump in the electron distribution has 
propagated to $x\approx 900$ for $S_3$ and the proton beam ahead of the
shock has started to thermalize by its interaction with the upstream plasma, 
as it is evidenced by the Fig. \ref{fig7}. An electron phase space hole 
doublett and proton phase space structures are visible. These structures 
have grown out of the phase space oscillation of the proton beams and the 
electron phase space hole at $x\approx 1250$ in Fig. \ref{fig6}(c). The 
proton distribution in Fig. \ref{fig7}(c) in $x\geq 1700$ reveals that a 
second shock is forming, which will thermalize the dense and fast beam of 
protons that expands out of the plasma 1 into the plasma 2. The spatially 
uniform electron distribution outside the interval occupied by the electron 
phase space holes changes only its thermal spread and density along $x$ 
and could be approximated by a Boltzmann-distribution. The electrons are 
not accelerated to high energies neither by the shocks nor by the other
phase space structures.

The expansion of the protons of plasma 1 in simulation 2 is captured by the 
movie 3. The colour scale corresponds to the 10-logarithmic number of CPs. 
The movie 3 evidences the formation of the shock and of its downstream 
region and it displays how the proton phase space hole and, subsequently, 
the secondary shock develop. The mean velocity of the upstream protons is 
modulated along $x$, which is probably a result of the same wave fields that
thermalized the electrons.

\begin{figure}
\begin{center}
\includegraphics[width=6.5cm]{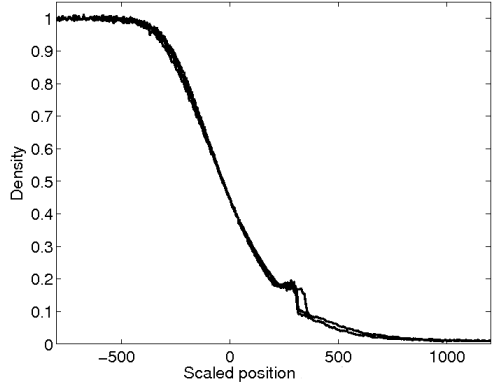}
\caption{The proton densities, normalized to $n_h$, as a function of the 
scaled position $x t_1 / t_j$, where $t_j$ corresponds to the snapshot $S_j$. 
The curves match, except within the downstream region of the shock at $200 
< xt_1/t_j < 400$ that is characterized by a constant density. The density 
doubles by the shock compression at $xt_1 / t_j \approx 350$.\label{fig8}}
\end{center}
\end{figure}

The proton distribution at $x\approx 0$ changes in time primarily due to the 
free motion of a proton $i$ with the speed $v_{x,i}$, which is displaced as 
$x_i = v_{x,i}t$. The phase space boundary between the plasma 1 and 2 is thus 
increasingly sheared. Further acceleration mechanisms are the drag of the 
protons by the thermally expanding electrons and the shock formation. Figure 
\ref{fig8} assesses their relative importance. The plasma density distribution 
should be invariant if the protons expand freely and if we scale the position 
$\propto x / t$. This is indeed the case and the proton density distributions 
for $S_1,S_2$ and $S_3$ match if we use the scaled positions, 
except at the shock and within its downstream region. The electron densities 
(not shown) closely follow those of the protons.
 
\begin{figure}
\begin{center}
\includegraphics[width=7cm]{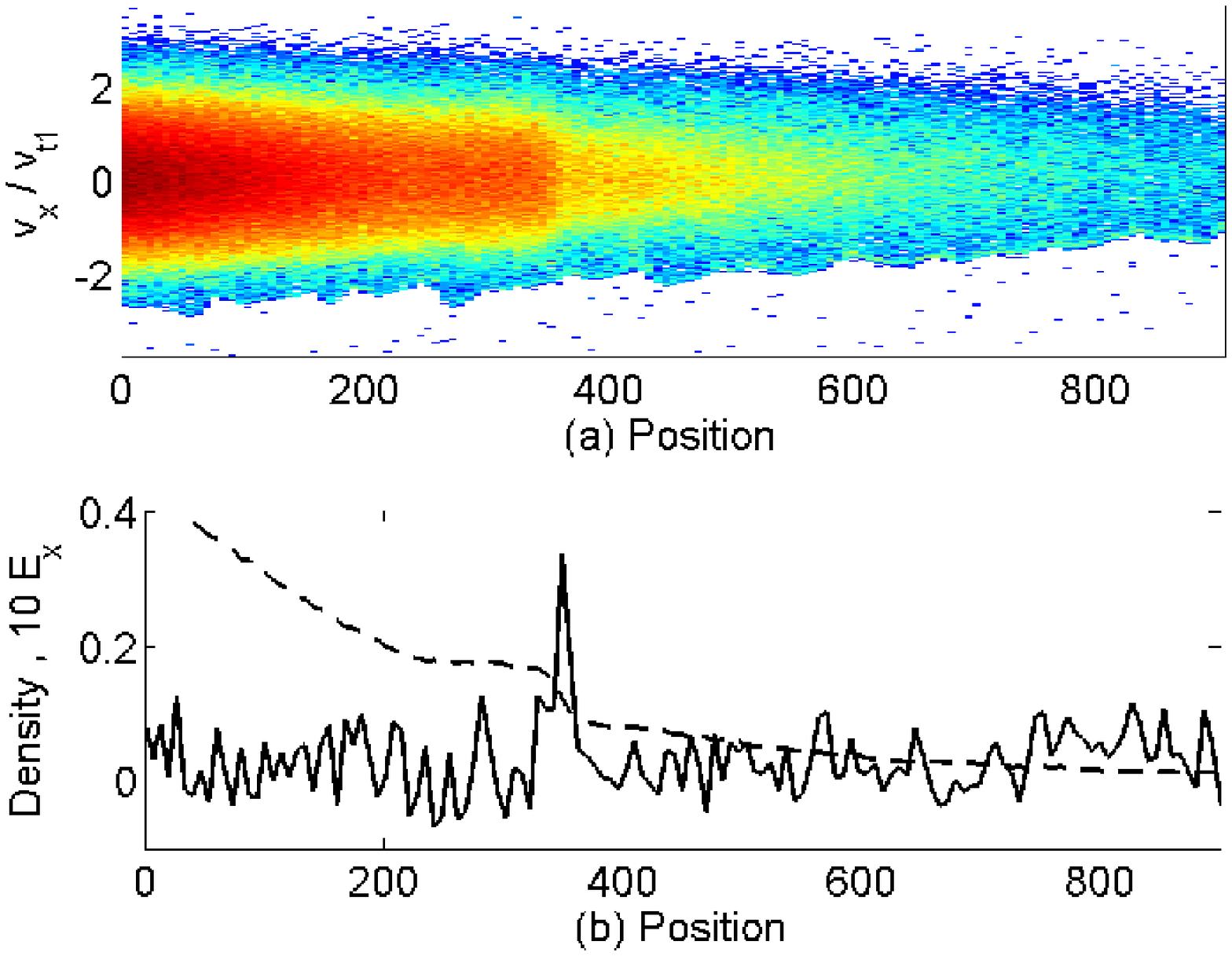}
\includegraphics[width=8cm]{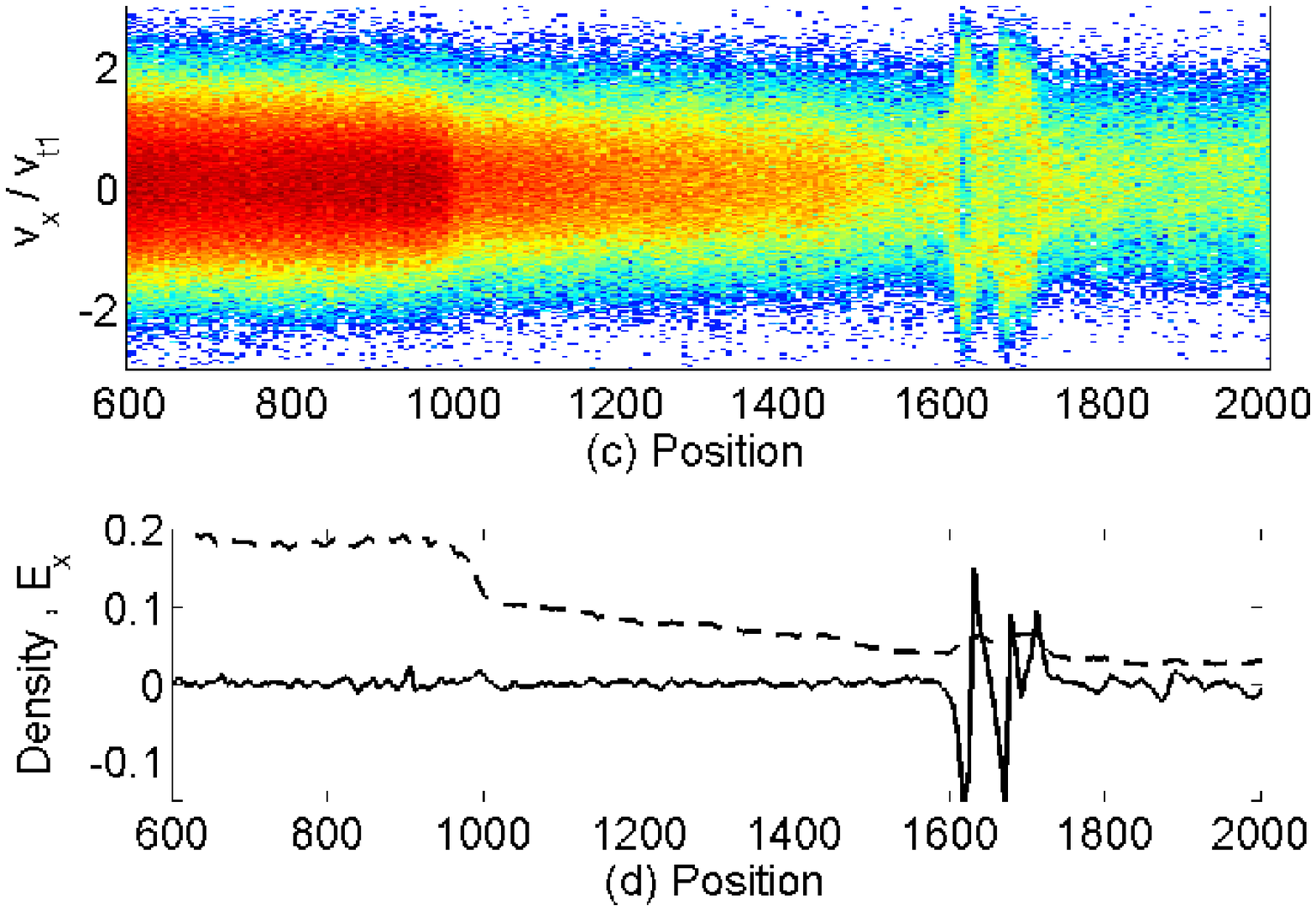}
\caption{(Colour online) The 10-logarithmic electron phase space 
distributions are shown in (a) for $S_1$ and (c) for $S_3$. The electron 
density (dashed curves) and the electrostatic field (solid curves) 
are displayed in (b) for $S_1$ and in (d) for $S_3$. The densities are
integrated and the electric fields averaged over 5 grid cells.\label{fig9}}
\end{center}
\end{figure}

Figure \ref{fig9} compares the electrostatic field with the electron
distributions for the snapshots $S_1$ and $S_3$.
An electric field peak at $x\approx 400$ coincides with the shock in the
snapshot $S_1$. The peak $E_x \approx 0.04$ and it confines the electrons 
to the left of the density jump by accelerating them into the negative $x$ 
direction. The electric field can be scaled to physical units with 
$n_M=10^{15}$ ${\rm cm}^{-3}$ and $v_{t1}=1.325 \times 10^7$ m/s to give 
$\approx 5 \times 10^6$ V/m. The electric field, which has been measured 
close to the shock in Ref. \cite{Romagnani}, is $\leq 2 \times 10^7$ V/m. 
The plasma density in the region, where the shock develops in the experiment, 
may be higher than $10^{15}$ ${\rm cm}^{-3}$. The electric field amplitudes 
associated with the shock are thus comparable. The noise levels in PIC 
simulations are typically higher than in a physical plasma, explaining the 
strength of the evenly spread noise in the simulation box, which is not 
observed to the same extent in the experiment. The electric field at the 
shock at $x\approx 10^3$ is at noise levels for $S_3$, while the phase 
space holes at $x\approx 1700$ give an electric field, which is exceeding 
that sustained by the shock for $S_1$.

\section{Discussion}

We have investigated the thermal expansion of a hot dense plasma into 
a cooler tenuous plasma. The thermal pressure of the hot plasma exceeded 
that of the cool plasma by the factor $10^4$. Our study has been motivated 
by the laser-plasma experiment in Ref. \cite{Romagnani}, which examined the
expansion of a hot and dense plasma into a tenuous ambient medium. Our 
initial conditions and the 1D geometry are, however, idealized and the 
simulation results can thus not be compared quantitatively to the 
experimental ones. The aim of our work has been to better understand the 
qualitative effects of the ambient medium on the plasma expansion. We have 
for this purpose compared our results with some of those in the related 
study in Ref. \cite{Mora2}, that considered the plasma expansion into a 
vacuum. There, the electron temperature exceeded that of the protons by a 
factor $10^3$, while we consider here the same temperature of electrons 
and protons.

Our results are summarized as follows. An electric field grows almost 
instantly at the boundary between both plasmas, because the ion expansion 
of the hot plasma is slower than the electron expansion. The electric
field forms irrespective of the ambient medium. It accelerates only the 
ions, if the plasma expands into a vacuum and it has a cusp in its spatial 
profile. The acceleration of the electrons of the ambient medium triggers 
in our simulation the formation of a double layer \cite{Ishiguro} with a 
smooth electric field profile. This double layer redistributes the momentum 
between the individual plasma species \cite{DL3}. A tenuous hot beam of 
electrons streams from the hot plasma into the cool plasma, while all the 
electrons of the cool plasma are dragged into the hot plasma. These beams 
thermalize through electrostatic two-stream instabilities, which equilibrate 
the electron temperatures of both plasmas on electron time scales. This rapid 
thermalization will cancel any significant proton acceleration by hot 
electrons already at the relatively low density of the ambient medium we 
have used. Proton acceleration is, however, still possible because a 
thermal pressure gradient is provided by the density jump. Most electrons 
merely follow after their thermalization the motion of the protons to 
conserve the quasi-neutrality of the plasma. They maintain their Maxwellian 
velocity distribution, which would not be the case for an expansion into
a vacuum \cite{Mora2}.

The protons at the front of the hot plasma are accelerated by the electric 
field of the double layer to about 5.5 times the proton thermal speed, while
the Maxwellian distribution is represented up to 3-4 times the proton thermal
speed. The expansion of the protons from the hot into the cool plasma is 
dominated by the free streaming of the fastest protons (diffusion). The 
effects of the ambient medium on the proton expansion are initially 
negligible. Eventually the interaction of the expanding and the ambient 
plasma results in the formation of shocks. We have observed one shock at 
the position, where the protons of both plasmas merge. This shock did not 
result in the acceleration of electrons or in the modification of their 
phase space distribution. 

The protons of the hot plasma expand farther than the position of this shock 
and they can interact with the protons of the cool plasma through ion beam 
instabilities. The interval, in which the protons of both plasmas co-exist, 
resembles qualitatively the upstream region of an electrostatic shock 
\cite{Forslund}. However, the density and the speed of the beam of expanding 
protons of the hot plasma are both higher than what we expect for a 
shock-reflected ion beam. We have observed in the simulation the growth of 
a phase space structure in the upstream proton distribution that gave rise 
to an electron phase space hole. The proton structure evolved into a second 
shock ahead of the primary one. The presence of multiple shocks has been 
observed experimentally \cite{SecondShock}, although there the second shock 
was radiation-driven and not beam-driven. 

\subsection{Acknowledgments}

The authors acknowledge the financial support by an EPSRC Science and 
Innovation award, by the visiting scientist programme of the Queen's 
University Belfast, by Vetenskapsr\aa det and by the Deutsche 
Forschungsgemeinschaft (Forschergruppe FOR1048). The HPC2N computer center 
has provided the computer time.

\end{document}